# Optical Rashba effect in a monolithic light-emitting perovskite metasurface

Jingyi Tian[1,2]†, Giorgio Adamo[1,2]†, Hailong Liu[3], Maciej Klein[1,2], Song Han[1],

Hong Liu[3], Cesare Soci[1,2]*

[1] Centre for Disruptive Photonic Technologies; TPI, SPMS, Nanyang Technological University, 21 Nanyang Link, Singapore 637371

[2] Division of Physics and Applied Physics, School of Physical and Mathematical Sciences, Nanyang Technological University, Singapore 637371

[3] Institute of Materials Research and Engineering, A*STAR (Agency for Science Technology and Research); 2 Fusionopolis Way, #08-03, Innovis, Singapore 138634

† These authors contributed equally to this work.

*Corresponding author. Email: csoci@ntu.edu.sg

**The Rashba effect[1], i.e., the splitting of electronic spin-polarized bands in the momentum space of a crystal with broken inversion symmetry, has enabled the realization of spin–orbitronic devices, in which spins are manipulated by spin-orbit coupling[2]. In optics, where the helicity of light polarization represents the spin degree of freedom for spin-momentum coupling[3], the optical Rashba effect is manifested by the splitting of optical states with opposite chirality in the momentum space[4-8]. Previous realizations of the optical Rashba effect relied on passive devices determining either the propagation direction of surface plasmons[4] or circularly polarized light into nanostructures[5], or the directional emission of polarized luminescence from metamaterials hybridized with light-emitting media[6-8]. Here we demonstrate an active device underpinned by the optical Rashba effect, in which a monolithic halide perovskite metasurface emits highly directional chiral photoluminescence. An all-dielectric metasurface design with broken in-plane inversion symmetry is directly embossed into the high refractive index, light-emitting perovskite film, yielding a degree of circular polarization of photoluminescence of 40% at room temperature - more than one order of magnitude greater than state of art chiral perovskites[9-11].**

Luminescence from light emitting media is generally omnidirectional, unpolarized, incoherent and oftentimes weak (Fig. 1a, panel i). This is unsuitable for applications that require high brightness, directionality, and polarization control. Metasurfaces consisting of a periodic arrangement of artificial atoms can create delocalized virtual optical states (VOS) with enhanced local density of states and deterministic energy (spectral) and momentum (spatial) distributions, thus offering a powerful tool to manipulate incoherent light emission[6-8,12].

Metasurface design dictates the nature of the VOS. For example, if the structure has in-plane inversion symmetry (e.g., square lattices of circular holes, pillars or gratings), some of the VOS with frequency overlapping with the continuum cannot radiate in free space due to the symmetry mismatch with plane waves, thereby remaining perfectly confined within the structure[13]. Such VOS are referred to as symmetry-protected bound states in the continuum (BICs). Due to their highly confined nature, BICs display extremely high-quality factors (theoretically infinite) and manifest themselves as far-field intensity and polarization singularities of integer topological charges in the luminescence emitted by the metasurface[14-17]. Conversely, if the in-plane inversion symmetry of the structure is broken (e.g., square lattices of triangular metamolecules), the integer charges of the BICs decompose into pairs of half-integer charges in the momentum space[18,19] (details in SI section IV). The resulting VOS ($|\pm\frac{1}{2}>$) are then associated to purely circularly polarized singularity points of far-field polarizations with nonzero intensity and opposite handedness (Fig. 1a, panel ii). By designing the metasurface so that the wavelengths of its VOS overlap with the luminescence spectrum of the light-emitting medium, emission is funneled into the radiative channels opened by the VOS, whose polarization and directivity can be freely and precisely engineered (Fig. 1b). This effectively realizes the optical analogue of the Rashba effect in condensed matter.

After the theoretical prediction of its occurrence in zinc blende and wurtzite crystal lattices[20,21], the Rashba effect has been observed in a variety of condensed-matter systems, including 2D semiconductors, surfaces of metals, heterostructures and topological insulators[1]. Recently, the discovery of the Rashba effect in halide perovskites has triggered an enormous interest for their potential application in spintronics[22-25]. Besides their unique electronic properties, halide perovskites are an emerging optical platform for all-dielectric metamaterials that combine light confinement at the nanoscale and strong light-matter interaction[26] with excellent radiative properties[27,28]. This owes to the unique combination of high luminescence quantum yield and compositionally tunable emission spectrum with a high refractive index (n>2). Early realizations of dielectric perovskite metamaterials have been primarily focused to the demonstration of luminescence enhancement[29,30], with a few works recently extending the concept to the control of polarization and spatial distribution of microlaser emission[15-17]. A hallmark of polarization control, emission of circularly polarized light of chosen helicity, has only been demonstrated by incorporating chiral ligands into the inorganic framework of the perovskites, yet with a degree of circular polarization (DOP) not exceeding ±3% at room temperature[10,11], which is impractically small for chiral emitting device applications. The combination of metamaterial design strategies with the unique optoelectronic properties of halide perovskites opens new avenues for the implementation of highly chiral emission sources underpinned by the optical Rashba effect.

For this proof of principle demonstration, we spin-cast a 170 nm thick film of methylammonium lead-iodide (MAPbI$_3$) perovskite (optical properties in SI section II) in which polycrystalline domains are randomly distributed. This results in uniformly distributed and unpolarized radiation, resembling isotropic media made of an ensemble of randomly oriented dipoles with unrelated initial phases and locations. The film was patterned by nanoimprint lithography (NIL) to realize a periodic array of triangular metamolecules (Fig. 1c) with induced VOS located above the light cone and below the diffraction limit, which offer enhanced radiation channels into the far field with well-defined distributions of polarization and in-plane wave vectors.

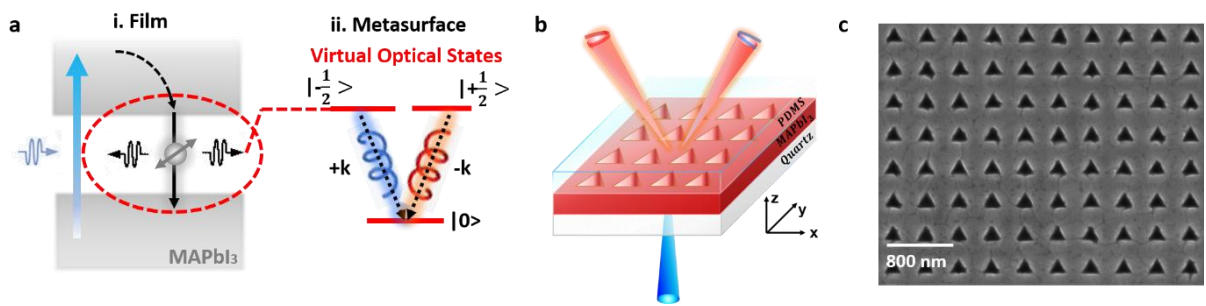

**Fig. 1 Perovskite metasurfaces for directional control of chiral luminescence** (a) Diagram of photoluminescence (PL) from MAPbI$_3$. (i) PL from a MAPbI$_3$ film due to near-band-edge transition resembles the emission from multiple dipoles, which is isotropic and unpolarized. (ii) PL from a MAPbI$_3$ metasurface with well-controlled far-field polarization distributions due to the splitting of virtual optical states (VOS) with opposite chirality ($|\pm\frac{1}{2}>$) in the momentum space. (b) Schematic of optical Rahsba effect in an all-dielectric in-plane inversion-symmetry-broken perovskite metasurface patterned with a square lattice of equilaterally triangular holes and sandwiched between a quartz (n=1.5) substrate and a PDMS (n=1.5) layer. The side length of the hole is 220 nm and the period is 400 nm. The metasurface is excited by blue laser (405 nm) from back side and the PL with opposite chirality is routed into opposite sectors of hemisphere. (c) Scanning electron microscope image of a $MAPbI_3$ metasurface made by nanoimprinting lithography. The scale bar is 800 nm.

The VOS of the metasurface can be identified by its optical band diagram. The calculated transverse electric (TE) polarized optical bands along the *x*-axis are illustrated in Fig 2a, where the wavevectors are linked to the far-field radiation angle $\theta$ of the VOS in the *x-z* plane, by the relation $k_x/k_0 = \sin\theta$. The TE2 band, highlighted in green, is a quadrupole state with low dispersion, located within the emission spectral range of $MAPbI_3$ around $\lambda = 800$ nm. The *z* component of the magnetic field, $H_z$, shows an asymmetric distribution of the TE2 mode in the *x-y* plane (inset in Fig. 2a), dictated by the broken in-plane inversion symmetry of the design. When TE2 is projected into the far-field as polarization vectors, $E_{xy}(\theta) = \left(E_x(\theta), E_y(\theta)\right)$, it is possible to calculate the polarization handedness of the electric field intensity as a function of the radiation angle. The resulting angular distribution of circularly polarized intensities with opposite handedness, $I_R(\theta)$ and $I_L(\theta)$, is strongly directional into different sectors of the far field ($\pm\theta$), as shown in Fig. 2b. The DOP of the radiated field (Fig. 2c) expressed as $DOP(\theta) = [I_R(\theta) - I_L(\theta)]/[I_R(\theta) + I_L(\theta)]$, reveals a quite rich far-field polarization distribution associated to the TE2 band. At $\theta = 0$ (normal direction), the mode polarization is linear along the *x* axis, here denoted by **H**. Moving just off normal in opposite directions along the *x*-axis, a pair of purely circularly polarized VOS of opposite handedness (DOP = $\pm$100%), denoted as **RC** and **LC**, is generated. Going further off the normal, the two circular polarizations first evolve into linear polarizations along the *y* axis, denoted by **V**, before switching handedness and reaching a $DOP_{sim} > \mp 60\%$. The evolution of the DOP can be mapped onto a Poincare sphere (inset of Fig. 2c) where it traces a vertical loop across the two poles representing circular polarizations **RC** and **LC**, and the equator corresponding to linear polarizations **H** and **V**. The angular separation between polarization states can be modified by either selecting different optical bands or adjusting the asymmetry factor of the metamolecule (Figs. S2 and S3).

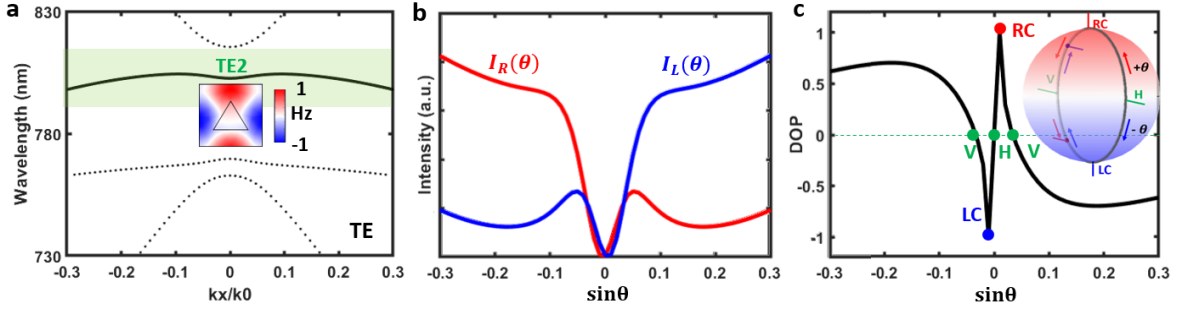

**Fig. 2 The virtual optical states of the metasurface.** (a) The VOS with TE polarized field in the metasurface and in-plane wavevectors along the *x*-axis. The optical band TE2 highlighted in green is a quadrupole state. The inset shows its field distribution within each unit cell on the *x-y* plane. (b) Distribution of far-field electric field intensity with different handedness from TE2 versus radiation angle $\theta$. (c) DOP of the radiated field from TE2 versus radiation angle $\theta$. The corresponding field vectors are mapped onto a Poincare Sphere, as a black vertical loop, to illustrate the full control of polarizations. The field is linearly polarized at $\theta = 0$ (**H**) and $sin\theta = \pm 0.035$ (**V**), purely circularly polarized at $sin\theta = \pm 0.005$ (**RC** and **LC**), and strongly circularly polarized at large radiation angles with DOP $> \pm 60\%$.

The calculated far-field intensity distribution and polarization maps of TE2 in two-dimensional momentum space are shown in Fig. 3a, where each position in the map corresponds to a far-field radiation direction defined by the $k_\| = (k_x, k_y)$. The polarizations across the radiated far-field vary from linear (**H**, **V**) to purely circular (**RC**, **LC**) with all intermediate polarization states, spanning the entire Poincare sphere[18]. Notably, circularly polarized radiation of different handedness is projected into opposite sectors of hemisphere ($\pm k_x$) in the far field (Fig. 3b and 3c). The calculated far-field DOP distribution of TE2 in the $k_\|$ space is shown in Fig. 3d. DOP is of opposite sign in the two $\pm k_x$ hemispheres, reaching its maximum values of $DOP_{sim} =$

±100% at radiation angles close to the normal, and $DOP_{sim} > \mp 60\%$ moving away from the center of the momentum space. The simulation results indicate that the VOS of the metasurface is expected to project distinct far-field polarizations into well separated directions.

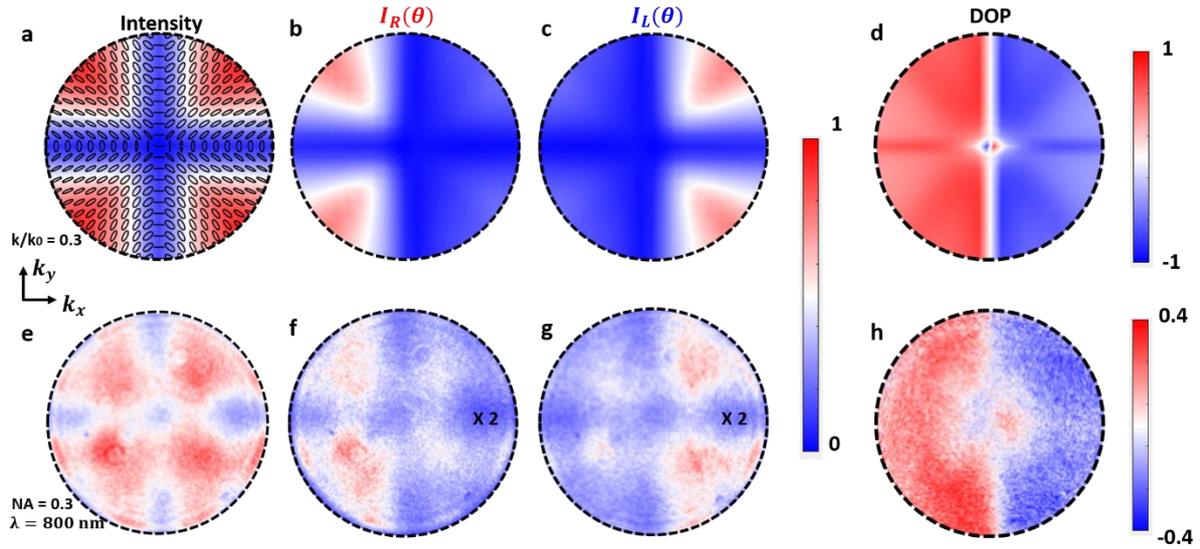

**Fig. 3 Optical Rashba effect of the light-emitting perovskite metasurface.** The calculated far-field (a) total electric field intensity distribution and polarization map, (b) right-handed polarized electric field intensity distribution, (c) left-handed polarized electric field intensity distribution and (d) DOP distribution of TE2 in the momentum space ($kx, ky$). The measured (e) total PL intensity distribution, (f) right-handed polarized PL intensity distribution, (g) left-handed polarized PL intensity distribution and (h) DOP of PL from the metasurface in the back focal plane captured by a CCD with a 10nm linewidth bandpass filter at 800 nm.

The spatially varying far-field polarizations of TE2 can be imprinted onto the MAPbI$_3$ luminescence, routing the emitted light of designed polarizations into different, predetermined directions. The angular distribution of the photoluminescence (PL) emitted by the MAPbI$_3$ metasurface was measured by back-focal plane imaging in an optical microscope (setup schematic in Fig. S4). The PL spectrum was filtered at $\lambda = 800$ nm to isolate the radiation coupled to TE2. The unpolarized emission was collected by an objective with numerical aperture (NA) of 0.3 and imaged by a CCD, using $\lambda/4$ waveplate and a linear polarizer to identify the chiral emission of opposite handedness. The measured intensity distributions of unpolarized, right and left circularly polarized PL and the DOP (Fig. 3e-h) across a solid angle 17.5° (NA = 0.3) are in extremely good agreement with the numerical predictions (Fig. 3a-d), which is the manifestation of optical Rashba effect. The experimental DOP reaches a remarkable value of $DOP_{exp} \sim \pm 40\%$, more than an order of magnitude improvement over previous chiral perovskite demonstrations[10,11]. Note that, due to the bandwidth of the bandpass filter used in the measurements, the collected PL also includes off-resonance radiation channels and VOS with opposite chirality adjacent to TE2, thus the $DOP_{exp}$ should only be considered as lower bound values.

Finally, not only the VOS spatially redistribute the chiral PL of the meatsurface, they also induce a significant Purcell enhancement of the emission intensity. This is shown by the comparison of normalized PL emission maps and spatially integrated emission intensity of the unpatterned $MAPbI_3$ film and the metasurface (Fig. 4). As expected, the PL from the unpatterned film around λ=800 nm is isotropic (Fig. 4a), while the metasurface shows spatial redistribution of the PL intensity with a maximum enhancement up to six-fold into specific directions (Fig. 4b). The spectral dependence of the spatially integrated PL intensity of the unpatterned $MAPbI_3$ film and the metasurface are shown in Fig. 4c, alongside the calculated enhancement, $\Gamma = I_{MS}/I_{Film}$. The enhancement reaches the maximum value of 4 at λ = 800 nm (the wavelength of the TE2 band), which is a manifestation of the Purcell effect. The PL enhancement at 810 nm and 780 nm are due to the Purcell effect from other VOS, i.e., TE1 and TM1 (Fig. S3).

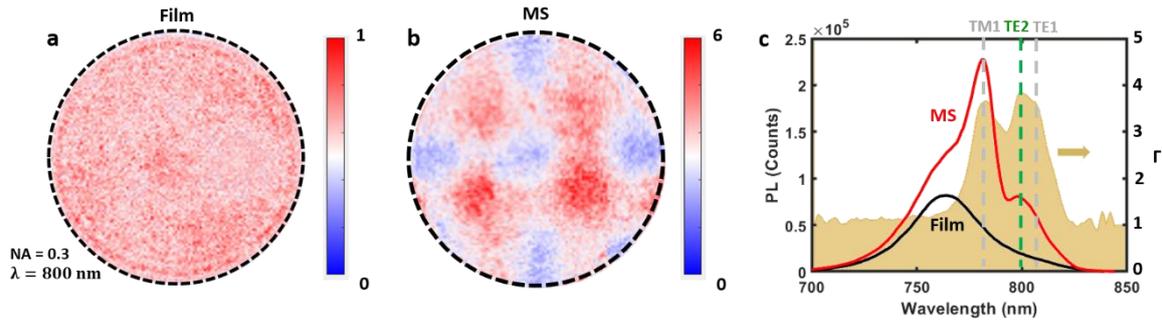

**Fig. 4 PL Enhancement of the metasurface.** The distribution of the PL intensity from (a) a bare MAPbI$_3$ film and (b) the metasurface captured by the 10 nm linewidth bandpass filter at 800 nm. The PL from the metausrface is normalized to that of the MAPbI$_3$ film. (c) The overall PL spectra of the MAPbI$_3$ metasurface and a bare film along with the PL enhancement due to the metasurface.

In summary, we demonstrated the optical Rashba effect in an active MAPbI$_3$ perovskite metasurface with broken in-plane inversion symmetry that allows generation, routing and enhancement of chiral photoluminescence. We experimentally achieved a DOP around 40% at room temperature, more than one order of magnitude larger than in state-of-the-art chiral perovskites, and a six-fold enhancement of photoluminescence intensity due to the concurrence of Purcell effect and a far-field intensity redistribution. The implementation of the optical Rashba effect in a monolithic light-emitting materials platform represents a true analogue of the electronic Rashba effect in condensed matter systems and may facilitate the development of planar light emitting nanodevices with potential applications in holography, bio-sensing and encoding, polarization-division multiplexing and quantum optics.

**Acknowledgements**

We acknowledge Jie Deng and Norman Soo Seng Ang for assistance with fabrication of the nanoimprint lithography mask. Research was supported by the A*STAR-AME programmatic fund on Nanoantenna Spatial Light Modulators for Next-Gen Display Technologies (Grant A18A7b0058), and the Singapore Ministry of Education (MOE2016-T3-1-006).

**Author contributions**

C.S., J.T. and G.A. conceived the idea. J.T. performed the numerical simulations and theoretical analysis with help from S. Han. J.T. and G.A designed the experiments, developed the back focal plane imaging setup and performed the measurements. G.A. performed the SEM measurement. M.K. synthesized the perovskite film. Hailong L. performed the nanoimprint lithography process under the supervision of Hong L. J.T., G.A. and C.S. drafted the manuscript and all authors contributed to the writing.

**Competing financial interests**

The authors declare no competing financial interests.

**Additional information**

**Supplementary information** is available in the online version of the paper.

**Data availability**

The authors declare that all data supporting the findings of this study are available within this article and its supplementary information and are openly available in NTU research data repository DR-NTU (Data) at https://doi.org/XXXXXX. Additional data related to this paper may be requested from the authors.

# Supporting Information

## Optical Rashba effect in a monolithic light-emitting perovskite metasurface

Jingyi Tian[1,2]†, Giorgio Adamo[1,2]†, Hailong Liu[3], Maciej Klein[1,2], Song Han[1],

Hong Liu[3], Cesare Soci[1,2]*

[1] Centre for Disruptive Photonic Technologies; TPI, SPMS, Nanyang Technological University, 21 Nanyang Link, Singapore 637371

[2] Division of Physics and Applied Physics, School of Physical and Mathematical Sciences, Nanyang Technological University, Singapore 637371

[3] Institute of Materials Research and Engineering, A*STAR (Agency for Science Technology and Research); 2 Fusionopolis Way, #08-03, Innovis, Singapore 138634

† These authors contributed equally to this work.

*Corresponding author. Email:* csoci@ntu.edu.sg

# I MATERIALS AND METHODS

## 1. Sample preparation

### (1) Film preparation

Quartz substrates are cleaned through immersion in the mixture of 2 mL of Hellmanex II (Hellma Analytics) and 200 mL of deionized (DI) water at 353 K for 10 min, subsequently rinsed with DI water and dried in nitrogen flow followed by oxygen plasma treatment. $CH_3NH_3I$ (Dyesol) and $PbI_2$ (99.99%, TCI) powders are added to anhydrous dimethylformamide (DMF, Sigma-Aldrich) and stirred for several hours at room temperature to form 1.2 M precursor solution (molar ratio 1:1) which is then filtered by a polyvinylidene fluoride (PVDF) syringe filter (0.45 μm) and heated on a hot plate at 373 K for one hour. Hot solution is spin-coated onto quartz substrates at 4900 rpm for 30 s, with dripping of toluene after 5 s of spinning time. Prepared $MAPbI_3$ films are finally annealed at 373 K for 15 min. Precursor solution preparation and film fabrication steps are carried out in $N_2$ filled glovebox.

### (2) Mold Fabrication and Thermal Nanoimprint Lithography Process

A negative tone resist (Hydrogen silsesquioxane, XR-1541-006) is spin-coated on a silicon substrate at a speed of 1500 rpm for 1 min. E-beam lithography is conducted to fabricate the triangle metasurface via ELS-7000 (Elionix Inc.) under an acceleration voltage of 100 kV and the dose of 7600 μC/cm$^2$. Inductively coupled plasma etching is employed as the following step to etch the Si substrate with a recipe of HBr (50 sccm) and $O_2$ (3 sccm) gases at 5 mTorr. Master molds are achieved after removing the HSQ mask in a buffered hydrofluoric acid. A nanoimprinter (Obducat NIL-60-SS-UV-Nano-imprinter) is used to transfer the metasurface from master mold to $MAPbI_3$ film at 30 bar and 90°C, and the imprinting time is optimized as 30 min. The imprinted sample is cooled down to 30°C and manually demolded from the master mold.

## 2. Numerical Simulations

The optical bands of perovskite metasurfaces are calculated using three-dimensional finite-element method (COMSOL). One unit cell, consisting of a triangular hole at the center (with index of air), is simulated when embedded in a homogeneous background (n = 1.5). Periodic boundary conditions are adopted in both x and y directions and perfectly matched layers (PML) along the z-direction are constructed. The optical bands and the corresponding far-field vector distributions in the momentum space is calculated with eigen-frequency solver by sweeping in-plane wavevectors.

# II OPTICAL CONSTANTS OF MAPBI3 AT ROOM TEMPERATURE

The optical properties of MAPbI$_3$ within the gain spectral region at room temperature from ellipsometry measurement are shown in Fig. S1. As shown in Fig. S1 (a), it exhibits relatively high refractive indices, allowing to effectively confine and modulate light at subwavelength scale.

In Fig. S1 (b), the absorption edge of MAPbI$_3$ at room temperature is at about 770 nm.

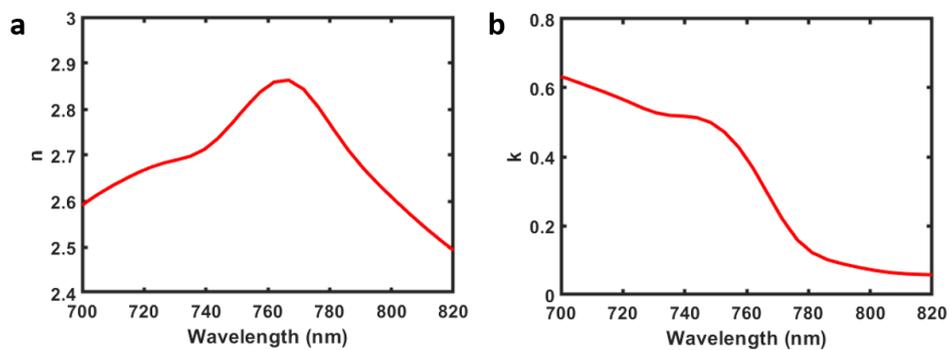

Figure S1(a) Real part and (b) imaginary part of refractive index of MAPbI$_3$ at room temperature.

# III CONTROLLING THE SEPARATING ANGLES OF OPPOSITE CIRCULARLY POLARIZED STATES

For a virtual optical state (VOS) with in-plane wave vector $k_{\parallel} = (kx, ky)$ above the light line and below the diffraction limit inside the metasurface, it generally couples to free space with the same $k_{\parallel}$. Its far-field distribution in $k_{\parallel}$ space can be described by projecting a 2D polarization vectors onto the *x-y* plane in terms of circular basis [S1],

$$d(k_{\parallel}) = d_+(k_{\parallel})e_+ + d_-(k_{\parallel})e_- \tag{S1}$$

Where $\mathbf{e}_{\pm} = (\hat{x} \pm i\hat{y})/\sqrt{2}$. In this basis, $d_+(k_{\parallel}) = 0$ or $d_-(k_{\parallel}) = 0$ corresponds to a circular polarized state (CPS). When both $d_+(k_{\parallel}) = 0$ and $d_-(k_{\parallel}) = 0$, the VOS turns into a Bound states in the continuum (BICs) with no emission into the far field. BICs carry conserved and quantized topological charges as polarization vortex centers. The topological charge ($q$) carried by BICs is defined as [S2]

$$q = \frac{1}{2\pi}\oint_C dk_{\parallel} \cdot \nabla_{k_{\parallel}}\varphi(k_{\parallel}), \quad q \in Z \tag{S2}$$

Where $\varphi(k)$ is the angle between long axis of polarization ellipse and *x*-axis. C denotes a closed loop around BIC counterclockwisely in the momentum space. $q$ indicates how many rounds the long axis of the polarization ellipse winds around the BIC, which must be an integer. By writing $d_{\pm}(k_{\parallel}) = |d_{\pm}|e^{i\alpha_{\pm}}$, we now define an integer[S1],

$$q_{\pm} = \frac{1}{2\pi}\oint_{C1} dk_{\parallel} \cdot \nabla_{k_{\parallel}}\alpha_{\pm}(k_{\parallel}), \quad q_{\pm} \in Z \tag{S3}$$

C1 denotes a closed loop around $d_{\pm}(k_{\parallel}) = 0$ counterclockwisely in the momentum space.

Since $\varphi(k) = \frac{1}{2}\arg(d_+^* d_-)$, $q$ and $q_{\pm}$ are related by[S2],

$$q = \frac{1}{2}(-q_+ + q_-) \tag{S4}$$

A BIC with $q = \pm 1$ can thus be interpreted as the superposition of two circularly polarized states with $(q_+, q_-) = \pm(-1, 1)$. This explains the phenomenon that by breaking the in-plane inversion symmetry, the BIC can split into two CPSs in the $k_{\parallel}$ space with the same topological charge $q$ (half integers) and opposite handedness $q_{\pm}$.

The separation of the two CPSs in the $k_{\parallel}$ space can be controlled by asymmetric parameters, which indicates to what extent the in-plane inversion symmetry has been broken. For example, in Fig. S2, while fixing the base and height of the trapezoid hole as $d = 220$ nm, the top dimension of the trapezoid hole is changed from 0 to 220 nm and denoted by $d1$. When $d1 = 220$ nm, the trapezoid hole becomes a cubic hole which sustains in-plane inversion symmetry and there is no splitting of the BIC (i.e., no CPSs). When $d1 = 0$ nm, the trapezoid hole transforms into a triangular hole as demonstrated in the manuscript.

We start by looking at the Degree of Circular Polarization (DOP) of VOS with TE polarized field in the metasurface and in-plane wavevectors along $x$-axis. In Fig. S2a, for band TE1, the location of the two CPSs (where DOP = $\pm 1$) changes from 0 to $sin\theta \sim \pm 0.025$ when d1 decreases from 220 nm to 10 nm. For band TE2, the separation of the two CPSs first increases from $sin\theta = 0$ to about 0.018 when $d1$ decreases from 220 nm to 140 nm and then they approach each other again to $sin\theta \sim 0.005$ when $d1$ is further reduced to 10 nm (Fig. 2b). The calculated polarization map and DOP distribution of TE2 in the $\boldsymbol{k}_{\parallel}$ space $(kx, ky)$ when $d1 = 140$ nm is illustrated in Fig. 2c.

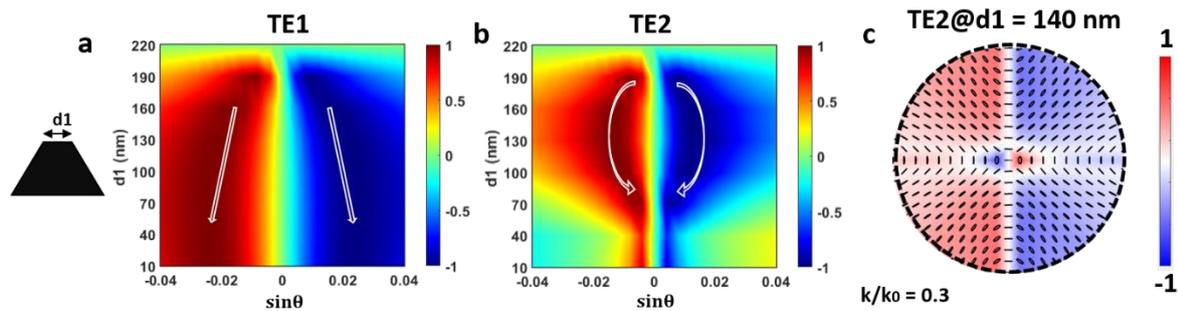

Figure S2 Calculated Degree of Circular Polarization (DOP) of VOS (a) TE1 and (b) TE2 versus radiation angle $\theta$. (c) The calculated polarization map and DOP distribution of TE2 in the $\boldsymbol{k}_{\parallel}$ space $(kx, ky)$ when $d1 = 140$ nm.

The separation of the CPSs in the $\boldsymbol{k}_{\parallel}$ space and far-field DOP distribution can also be controlled by working at different VOSs. For example, Fig. S3 illustrates two representative VOSs.

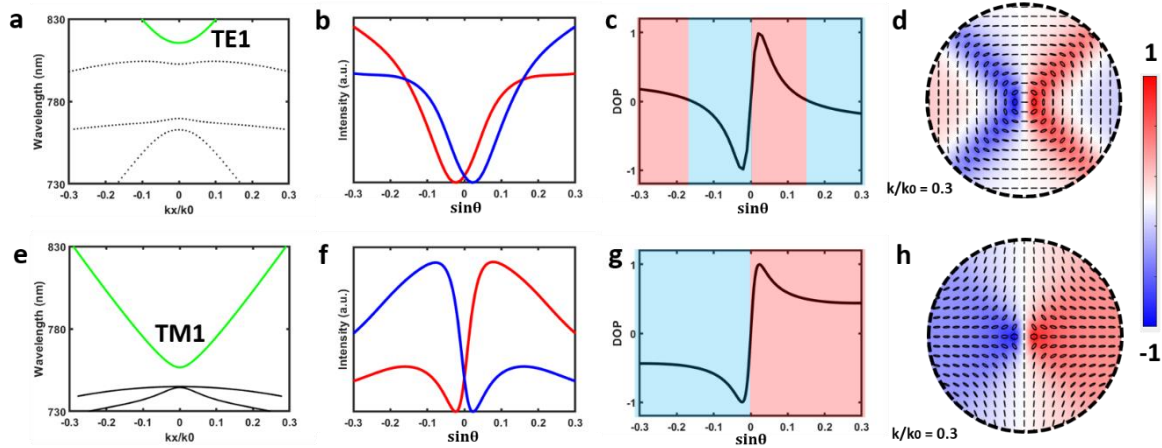

Figure S3 (a) The VOSs with TE polarized field in the metasurface and in-plane wavevectors along *x*-axis. The optical band TE1 highlighted in green is a magnetic dipole state. (b) Distribution of far-field electric field intensity with different handedness from TE1 versus radiation angle $\theta$. (c) DOP of the radiated field from TE1 versus radiation angle $\theta$. Red and blue regions denote right-handed and left-handed polarized light, respectively. (d) The calculated polarization map and DOP distribution of TE1 in the $\boldsymbol{k}_{\parallel}$ space $(kx, ky)$. (e) The VOSs with TM polarized field in the metasurface and in-plane wavevectors along *x*-axis. The optical band TM1 highlighted in green is a quadrupole state. (f) Distribution of far-field electric field intensity with different handedness from TM1 versus radiation angle $\theta$. (g) DOP of the radiated field from TM1 versus radiation angle $\theta$. Red and blue regions denote right-handed and left-handed polarized light, respectively. (h) The calculated polarization map and DOP distribution of TM1 in the $\boldsymbol{k}_{\parallel}$ space $(kx, ky)$.

## IV OPTICAL CHARACTERIZATION

**Polarimetric back focal plane imaging**

To precisely track the far-field distribution of the photoluminescence (PL) from the metasurface. Instead of using a traditional ellipsometry technique, polarimetric back focal plane imaging is conducted here [S3]. The setup is shown in Fig. S4. The metasurface is excited by a blue laser from back side through a NA = 0.1 condensor and the far-field distribution of PL intensity with emission angle smaller than the NA (NA = 0.3) of the collecting objective can be recorded on the 2D CCD arrays without scanning of different emission angles. In this setup, the back focal plane of the collecting objective is imaged rather than the sample. Each position on the CCD exclusively corresponds to a far-field radiation direction defined by $(kx, ky)$, where $\frac{k_x}{k_0} = \sin\theta$ and $\frac{k_y}{k_0} = \sin\varphi$. A quarter wave plate in combination of a linear polarizer is adopted to resolve the handedness of the PL. A bandpass filter with a 10nm linewidth at 800 nm is used to filter out the excitation laser and study specific emission wavelength at CCD.

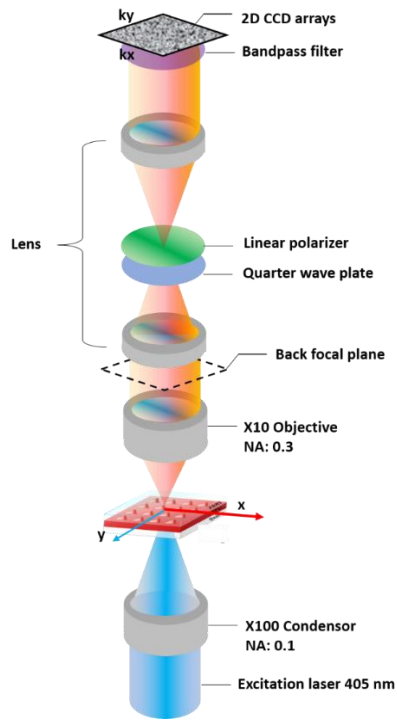

Figure. S4 Experimental setup for polarimetric back focal plane imaging

# V MEASURED FAR-FIELD DISTRIBUTION OF MAIN AXIS OF POLARIZATION ELLIPSE FROM TE2

The calculated far-field intensity distribution and polarization map of TE2 (which is mainly discussed in the manuscript) in the $\mathbf{k}_{\parallel}$ space $(kx, ky)$ is shown in Fig. S5a. To study the PL emission under the influence of TE2, angle-resolved PL mapping are performed by back-focal plane imaging with a collecting objective of NA = 0.3 and captured by a CCD with a linear polarizer and a 10nm linewidth bandpass filter at 800 nm (Fig. S5b-e), whose intensity distribution shows good correspondence to the numerical prediction of the main axis of polarization ellipse in Fig. S5a.

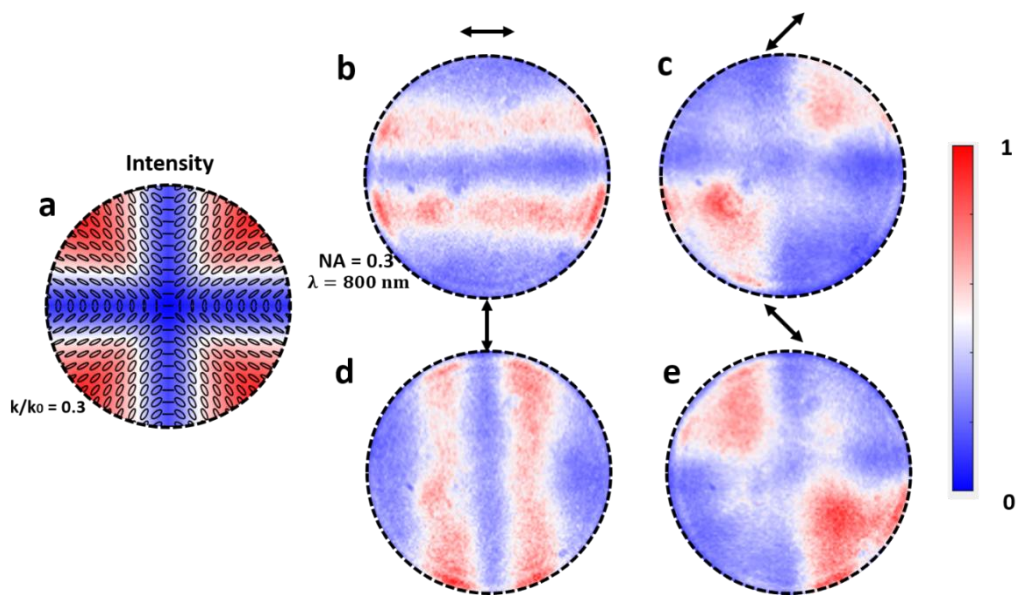

Figure S5 (a) The calculated far-field total electric field intensity distribution and polarization map of TE2 in the momentum space $(kx, ky)$. The measured PL intensity distribution with a linear polarizer at an angle of (b) 0° (c) 45° (d) 90° and (e) 135° with respect to the x-axis in the back focal plane captured by a CCD with a 10nm linewidth bandpass filter at 800 nm.